# A generalized Price equation for fuzzy set-mappings

*Matthias Borgstede**

*Corresponding author:

University of Bamberg

Markusplatz 3

D-96047 Bamberg

matthias.borgstede@uni-bamberg.de

# A generalized Price equation for fuzzy set-mappings

## Abstract

The Price equation provides a formal account of selection building on a right-total mapping between two classes of individuals, that is usually interpreted as a parent-offspring relation. This paper presents a new formulation of the Price equation in terms of fuzzy set-mappings to account for structures where the targets of selection may vary in the degree to which they belong to the classes of “parents” and “offspring,” and in the degree to which these two classes of individuals are related. The fuzzy set formulation widens the scope of the Price equation such that it equally applies to natural selection, cultural selection, operant selection, and even selection in physical systems.

## 1 Introduction

The Price equation provides an abstract description of selection on an arbitrary character value by partitioning the change in mean character value from a parent population to an offspring population, such that the amount of selection may be quantified by the covariance between the character value and evolutionary fitness (Price, 1970, 1972). The strength of the Price equation is that it provides a general framework for the theoretical analysis of evolutionary processes (Luque, 2017), such as kin selection (Frank, 1997), multilevel selection (Gardner, 2015), selection in class-structured populations (Grafen, 2015), and selection in uncertain environments (Grafen, 2000). Although originally introduced to account for natural selection on gene frequencies, the Price equation has been claimed to apply to other kinds of selection, as well (Frank, 2018; Price, 1970, 1995). Price-like partitionings have been proposed for several domains where selection may be effective, including cultural change (El Mouden et al., 2014), neural plasticity (Fernando et al., 2012), or operant learning (Baum, 2017). Some approaches even link different kinds of selection in one and the same model (Aguilar & Akçay, 2018; Borgstede & Luque, 2021). However, the scope of the Price equation remains limited by the structural assumptions from which it is derived.

In the most general case, the Price equation builds on two disjunct sets, $A$ ("parents") and $B$ ("offspring") and a left-unique, right-total relation $R \subseteq A \times B$ ("parentage") that connects every element in $B$ to exactly one element in $A$ (i.e., each offspring has one parent). Furthermore, the structure contains a real-valued function $z: A \cup B \mapsto \mathbb{R}$ (i.e., a quantitative character value). For all elements $i$ in $A$ the corresponding character value is denoted $z_i$. Furthermore, fitness $w_i$ is defined as the relative contribution of these elements to $B$ via the relation $R$ (i.e., parental fitness equals the number of offspring divided by the size of the offspring population). Given these definitions, the Price equation states that the change in average character value from $A$ to $B$ ($\Delta\bar{z}$) is exactly equal to the covariance between parental character value and parental fitness plus the expected value of the fitness-weighted deviation in character value between parents and offspring ($\Delta z_i$):

$$\bar{w}\Delta\bar{z} = \mathrm{Cov}(w_i, z_i) + \mathrm{E}(w_i\, \Delta z_i) \quad (1.)$$

Whenever a structure fulfils the above criteria, the Price equation provides a valid partitioning between selection and non-selection sources of character value change.[1] However, in domains other than evolutionary biology, an application of the Price equation is not always straightforward. For example, when applied to operant learning, it is by no means clear what the elements of the two sets should be and how they might be connected by a left-unique, right-total relation (Borgstede & Eggert, 2021). A similar problem arises in the context of cultural selection, where it may be difficult to conceptualize "parentage" in a way that is consistent with the structural requirements of the Price equation since cultural transmission is not unidirectional (i.e., an individual may be a cultural "parent" to some other individuals and be a cultural "offspring" of the same individuals simultaneously). Although some of these problems have been formally addressed in previous work, there is so far no adaptation of the Price equation that solves these problems on a general level.

[1] Proofs of the Price equation can be found in Luque (2017), Gardner (2020) and elsewhere.

This paper serves the purpose to provide a formal framework that allows to apply the Price equation to a much broader class of structures using fuzzy set-mappings instead of classical (crisp) set-mappings (Zadeh, 1965). In the following, I will provide an explicit description of a (very general) class of structures (which includes the classical set-mapping as a special case), and derive a corresponding Price-like partitioning of selection and non-selection sources of change. Furthermore, I will illustrate the scope of the derived fuzzy Price equation by sketching how it can be applied to describe selection processes in biology, physics, and social science.

# 2 Calculation

## 2.1 A note on vocabulary

Due to the wider scope of the partitioning to be derived below, I shall depart from the biologically inspired vocabulary used in most publications on the Price equation. Instead of reproducing individuals I will merely speak of different *measurements* of a quantitative character value $z$. Measurements may be obtained from different individuals or groups, but also from different contexts or points in time within the life of a single individual. In fact, the formulation is so general, that measurements need not even be associated with living individuals at all but may include technical systems, organizations or even abstract entities like thoughts or ideas. In line with this more abstract vocabulary, I shall not speak of parents and offspring, but of *sources* and *outcomes*. Finally, the connection between the two classes of measurements will be called a *recurrence* relation, which naturally includes biological reproduction, but also non-genetic mechanisms of transmission like imitation or instruction. Furthermore, the concept of recurrence also covers unobserved processes within individuals that might mediate the effect of one measurement on another, such as memory traces or internal representations.

## 2.2 Notation and definitions

Let $M$ be the domain set of measurements, on which we define two fuzzy sets, $A$ and $B$, standing for source and outcome observations. In contrast to classical ("crisp") sets, fuzzy sets do not necessarily have sharp boundaries, such that the question whether a measurement is an element of $A$ or $B$ is a matter of degree. In other words, each measurement may belong to either $A$ or $B$ or both to varying degrees. Formally, the degree to which each measurement belongs to the two fuzzy sets is modeled via two real-valued membership functions $m_i: M \mapsto [0,1]$ and $m_j: M \mapsto [0,1]$, respectively. These membership functions indicate the absolute degrees to which each measurement in $M$ is a source or an outcome (with 0 meaning no membership and 1 meaning full membership). Note that I use the indices $i$ and $j$ to distinguish the corresponding membership functions merely to remain consistent with the usual notation for parent and offspring populations, not because they range over different domains. To facilitate notation, I further introduce *relative* degrees of membership $\varphi_i = m_i / \sum m_i$ and $\varphi_j = m_j / \sum m_j$, such that $\sum \varphi_i = \sum \varphi_j = 1$. The expected values of the measurements in $A$ and $B$ for a quantitative character value, $z: M \mapsto \mathbb{R}$, are defined as weighted averages over the domain set $M$ using the corresponding membership degrees $\varphi_i$ and $\varphi_j$ as weighting factors, such that $\mathrm{E}(z_i) = \bar{z}_i = \sum \varphi_i z_i$ and $\mathrm{E}(z_j) = \bar{z}_j = \sum \varphi_j z_j$. The mean difference between outcome and source character values is further defined $\Delta\bar{z} = \bar{z}_j - \bar{z}_i$.

The recurrence of outcome measurements with regard to their sources is expressed by a fuzzy relation $R$ indicating recurrence, with membership function $r_{ij}: M \times M \mapsto [0,1]$. This membership function indicates the absolute degree to which to measurements are related (with 0 meaning they are completely unrelated and 1 meaning that they are fully related). I further define *relative* recurrence weights $\gamma_{ij} = r_{ij} / \sum \varphi_i \sum \varphi_j \, r_{ij}$, such that $\sum \varphi_i \sum \varphi_j \, \gamma_{ij} = 1$. Since every recurrence relation weight $\gamma_{ij}$ connects exactly one past observation to one present observation, it further holds that $\sum \varphi_i \sum \varphi_j \, \gamma_{ij} = \sum \varphi_j \sum \varphi_i \, \gamma_{ij}$. The recurrence relation allows for the prediction of each outcome value $z_j$ from the totality of contributions from all other measurements, with $\Delta z_j$ being the deviation between the predicted value of $z_j$ from its actual value: $z_j = \sum \varphi_i \gamma_{ij} z_i + \Delta z_j$. Thus, the recurrence relation captures the predictive value of source measurements to outcome measurements.

Finally, the weighted sum, $w_i$, of a source measurement's recurrence is called *fitness* and is defined as: $w_i = \sum \varphi_j \gamma_{ij}$. Consequently, the fitness of a source measurement is actually nothing but its average recurrence with regard to the outcome measurements. Following the above definition of expected values for the fuzzy sets $A$ and $B$, mean source fitness, $\overline{w}_i$, is given by $\overline{w}_i = \mathrm{E}(w_i) = \sum \varphi_i w_i$, which according to the scaling of the recurrence weights, $\gamma_{ij}$, is equal to 1.

## 2.3 Derivation of a generalized Price equation for fuzzy set-mappings

The above definitions will now be used to derive a generalized Price equation that includes the classical version (Equation 1) as a special case but allows for a much wider range of applications. We start with the definition of the mean change in character value from source to outcome measurements:

$$\Delta \bar{z} = \bar{z}_j - \bar{z}_i \qquad (2.)$$

$\bar{z}_j$ can be obtained by taking the expectation over all predicted outcome values. Hence,

$$\bar{z}_j = \sum \varphi_j \sum \varphi_i \gamma_{ij} z_i + \sum \varphi_j \, \Delta z_j \qquad (3.)$$

Since $\sum \varphi_i \sum \varphi_j \, \gamma_{ij} = \sum \varphi_j \sum \varphi_i \, \gamma_{ij}$, we can equivalently express $\bar{z}_j$ as:

$$\bar{z}_j = \sum \varphi_i \sum \varphi_j \gamma_{ij} z_i + \sum \varphi_j \, \Delta z_j \qquad (4.)$$

Mean character value in source measurements is given by $\bar{z}_i = \sum \varphi_i z_i$. Therefore, the mean change in character value is:

$$\Delta \bar{z} = \sum \varphi_i \sum \varphi_j \gamma_{ij} z_i + \sum \varphi_j \, \Delta z_j - \sum \varphi_i z_i \qquad (5.)$$

Since, by definition, $\sum \varphi_j \gamma_{ij} = w_i$, the first term of the right-hand side of the equation is $\sum \varphi_i \sum \varphi_j \gamma_{ij} z_i = \mathrm{E}(w_i, z_i)$. Furthermore, the second term of the right-hand side of the equation is $\sum \varphi_j \, \Delta z_j = \mathrm{E}(\Delta z_j)$ and the last term is $\sum \varphi_i z_i = \mathrm{E}(z_i)$. Furthermore, note that, due to the scaling of the fuzzy relation membership degrees, $\mathrm{E}(w_i) = 1$. Hence, we can re-write the mean change in character value as:

$$\Delta\bar{z} = \mathrm{E}(w_i z_i) - \mathrm{E}(w_i)\mathrm{E}(z_i) + \mathrm{E}(\Delta z_j) \qquad (6.)$$

Applying the standard definition of covariance, we get:

$$\Delta\bar{z} = \mathrm{Cov}(w_i, z_i) + \mathrm{E}(\Delta z_j) \qquad (7.)$$

This partitioning closely resembles the original Price equation, with the exception that due to the scaling of fitness, there is no need for weighting by average fitness on the left-hand side. Note also that $\mathrm{E}(\Delta z_j)$ is the expectation over outcome values, $z_j$, whereas the covariance term is calculated for source values $w_i$ and $z_i$. The partitioning holds for any domain set with arbitrary membership degrees $\varphi_i, \varphi_j$ (as long as there is at least one non-zero membership degree, each). The membership degrees may express anything that might justify a partitioning of observations into two different kinds. Similarly, the recurrence relation may express any kind of link between the observations with regard to the character value and may take arbitrary values $\gamma_{ij}$.

# 3 Application

In this section, I will present several applications of Equation 7 to illustrate the flexibility and generality of the fuzzy Price equation. All examples follow the same basic rationale. First, the domain of measurements is specified and assigned to the two fuzzy sets $A$ and $B$ by means of their absolute set membership degrees (i.e., the degree to which each measurement acts as a source and an outcome, respectively). Second, the absolute membership degrees of the recurrence relation $R$ are derived from a theoretically informed model of the mechanisms underlying the (partial) recurrence of measurements. The corresponding scaled relative weights can then be used to calculate the fitness for each measurement, alongside the corresponding covariance term using Equation 7.

## 3.1 Genetical Selection

The fuzzy set-mapping underlying Equation 7 can be used to retrieve the *classical Price equation* as a special case. In this application, the domain of observations is given by the measurements of a genetically transmitted trait $z$ for all individuals in a parent population and all individuals in an offspring population that together constitute the domain $M$. The parent and offspring populations are crisp sets that are identified by their characteristic functions, respectively. In other words, all parents have absolute membership degrees $m_i = 1$ and $m_j = 0$, whereas all offspring have absolute membership degrees $m_i = 0$ and $m_j = 1$. The relative membership degrees are obtained by dividing these values by the absolute number of parents and the absolute number of offspring, respectively. The recurrence relation $R$ is given by a crisp subset of $M \times M$, such that the absolute degree of relatedness $r_{ij}$ equals 1 if and only if individual $j$ is an offspring of individual $i$ and 0 if it is not. The relative recurrence weights are calculated by dividing these degrees of relatedness by the average number of offspring in the parent population.

This crisp partitioning captures the classical case where each offspring is related to exactly one parent. However, the fuzzy membership degrees allow for arbitrary weighting factors of parents and offspring. Therefore, individuals may be weighted by their long-term contribution to the future population by adjusting the membership degrees $m_i$ and $m_j$ to the respective parental and offspring reproductive values. The resulting partitioning would yield an average change in $z$ that is weighted by the parental reproductive values. Accordingly, the fitness values $w_i$ would be calculated based on the number of offspring, weighted by the corresponding offspring reproductive values. The fuzzy Price equation thus provides a straightforward way to incorporate variation between individuals, much alike to *class-structured* versions of the Price equation (e.g., Borgstede, 2024; Grafen, 2015; Lion, 2018; Taylor, 1990).

Furthermore, since the fuzzy relation $R$ is neither required to be left-unique, nor right-total, it readily generalizes to *sexually reproducing species* where each offspring has two (or possibly more) parents. To model sexual reproduction, the absolute degree of relatedness $r_{ij}$ needs to be adjusted to account for the relative contributions of the two parents to the offspring's genotype. In many species (such as mammals), both parents contribute the same proportion to their offspring's genotype. For two equally contributing parents the corresponding weights would be $r_{ij} = 0.5$ if and only if individual $j$ is an offspring of individual $i$ and $r_{ij} = 0$ if it is not. For *haplodiploid species* (such as honeybees), the absolute relatedness degree between parents and offspring would be $r_{ij} = 0.5$ for daughters (because females receive half of their genome form their mother and half from their father). Regarding sons, the absolute relatedness degrees would be $r_{ij} = 1$ for their mother and $r_{ij} = 0$ for their father (because males receive their complete genome from their mother).

## 3.2 Cultural Selection

The fuzzy Price equation can also be applied to *cultural selection,* where behavior is passed on non-genetically on from one individual to another (e.g., by imitation or by instruction). The evolving trait $z$ may be any quantitative behavior, such as relative time spent at an activity or the frequency of engaging in a certain behavior per time unit. Different degrees of influence between individuals may be incorporated into the fuzzy recurrence relation $R$ by assigning higher degrees of absolute relatedness $r_{ij}$ if a person $i$ has a strong influence on the behavior of a person $j$, and lower degrees if they have a smaller influence (although it will often be challenging to measure the amount of influence in practice). Cultural ancestry can further be modeled by choosing appropriate absolute membership degrees $m_i$ and $m_j$ to express how much an individual belongs to the set of sources ("influencers") and to the set of outcomes ("followers") regarding the behavior $z$ . In the simplest case, we may use crisp membership functions to divide the population into two disjunct sets (like in the classical Price equation). However, Equation 7 also holds if the two sets are overlapping.

## 3.3 Operant Selection

In addition to selection due to social interactions, behavior may change due to *operant selection*, which is traditionally called "reinforcement." In this case, the domain of measurements is given by repeated measurements of $z$ in one and the same individual. Again, $z$ can be any quantitative behavior and the repeated measurements can be thought of as behavioral observations taken under comparable circumstances (i.e., given the same contextual cues as in an operant conditioning experiment). The observed behaviors in a test trial are treated as outcomes by assigning corresponding crisp membership degrees (i.e., $m_j = 1$ if a measurement occurred during the test trial and $m_j = 0$ if it occurred before the test trial). Since the influence of past events decays over time, the membership degrees $m_i$ of set $A$ (sources) are modeled according to a strictly monotone decreasing function (e.g., exponential decay). To capture the effects of reinforcement, the absolute relatedness degrees $r_{ij}$ are chosen such that they are proportional to the amount of reinforcement associated with measurement $i$. Since reinforcement cannot influence past events, the absolute relatedness degrees are set to zero for all measurements $j$ preceding measurement $i$. The resulting covariance term in Equation 7 captures the amount of change in average individual behavior due to reinforcement.

## 3.4 Physical Selection

The above examples are all considered with measurements obtained from living organisms. However, Equation 7 may equally well be applied to describe *physical selection* processes occurring in nonliving matter. As an illustrative example, consider the energy transfer occurring at the surface boundary between a liquid and a gas. The measurements $z$ are the kinetic energies of the liquid particles taken at two different points in time, $t_1$ and $t_2$. The corresponding membership degrees to the source set and the outcome set are modeled according to a crisp distinction between $t_1$ and $t_2$ using absolute membership degrees 0 and 1 (like in the classical Price equation). Since the sum of all liquid particles' kinetic energies is proportional to the temperature of the liquid, the change in mean trait value (i.e., $\Delta\bar{z}$ ) is proportional to the surface's heat (i.e., the energy transfer between the liquid and the gas). The recurrence relation $R$ is defined such that the absolute relatedness degrees $r_{ij}$ express the contribution of a particle $i$ to the kinetic energy of a particle $j$ (including particle $i$ itself). Consequently, the fitness of a particle equals the total amount of kinetic energy transfer from that particle to its surrounding, and the sum over all $r_{ij}$ equals the total kinetic energy of the system. Given this choice of recurrence weights, the fuzzy Price equation would yield an average change of energy of $\Delta\bar{z} = 0$ for a closed system, which is equivalent to the law of *conservation of energy*. However, at the surface of the liquid, the particles do not only contribute to the internal energy of the liquid, but also to the internal energy of the surrounding gas due to evaporation. Evaporation occurs when the kinetic energy of a liquid particle at the surface is so high that it leaves the system (i.e., it moves from the liquid to the surrounding gas). If a particle leaves the system, it transfers all of its kinetic energy to the surrounding gas, thereby decreasing the total kinetic energy of the liquid and, consequently, its temperature. Since the probability that a liquid particle leaves the system is higher for particles with a higher kinetic energy, there will be a negative covariance between the particle's kinetic energy and its fitness. Evaporation therefore constitutes a genuine selection process that can be described by Equation 7.

# 4 Conclusion

I have derived a Price-like partitioning that holds in a much broader class of structures than the ones that are already available. It includes the original set-mapping between two disjunct populations (parents and offspring) as a special case. Above that, the partitioning also accounts for cultural and operant selection and thus unifies the existing domain-specific formulations within a single formalism. In fact, the proposed generalized Price equation holds for any system where one (possibly fuzzy) set of measurements is used to predict a different (also, possibly fuzzy) set of measurements, such as repeated observations within one individual that are divided into present and past observations, the effects of knowledge transfer from teachers to students, and even partially overlapping sets of cultural ancestors and cultural descendants. There are no restrictions on the recurrence relation, either. For example, recurrence may be mediated by genetic inheritance between parents and offspring, by instruction between teachers and students, by memory traces within one individual, or by imitation between the members of an arbitrary large group. Furthermore, the possible applications of the fuzzy Price equation are not limited to the behavior of living organisms but also include physical processes, such as thermodynamic heat transfer.

The wide scope of possible applications presented above makes the fuzzy Price equation a promising candidate for the foundation of a general theory of selection, as envisioned by Price (1995). Nonetheless, the fuzzy Prize equation still has some limitations. Most importantly, in its current form, it does not link different selection processes acting on the same domain. For example, cultural and operant selection are most certainly not independent sources of behavior change, since social consequences (such as others adopting one's own behavior) may themselves have reinforcing effects. On the other hand, the fact that selection in different domains may be described by the same abstract principle does not imply identical underlying mechanisms. For example, although heat transfer and operant learning can both be considered to be selection processes, they are certainly realized by completely different mechanisms. In other words, the Price equation tells us whether selection occurs or not, but not how it is realized. However, it can guide the construction of more specific selection models as an overarching theoretical principle. Furthermore, providing a shared formal vocabulary, the generalized Price equation presented in this paper facilitates the comparison and theoretical integration within and across domains (cf. Baravalle et al., 2024). The partitioning of observed change into selection and non-selection components itself is not an empirical hypothesis. Instead, the Price equation is a mathematical identity and, thus, true by definition. However, recognition of covariance is of great theoretical value in that it points to the essential characteristics of selection processes and may thus guide the construction of specific models of selection in such diverse fields as biology, sociology, psychology, and even physics.

# 5 Funding

This research did not receive any specific grant from funding agencies in the public, commercial, or not-for-profit sectors.

# 6 Conflict of interest

The author declares that the research was conducted in the absence of any commercial or financial relationships that could be construed as a potential conflict of interest.

# 7 Data Availability

This research has no associated data.